\documentclass[pra,nofootinbib]{revtex4-2}

 \usepackage{amssymb} \usepackage{graphicx}
\usepackage[utf8]{inputenc}
\usepackage[english]{babel} 
\usepackage{amsthm}
 \usepackage{amsmath}
 \usepackage{xcolor}
 \usepackage[normalem]{ulem}

\begin{document}
 \title{Spinor bilinears and Killing-Yano forms in generalized geometry}

\author{\"{O}zg\"{u}r A\c{c}{\i}k}
\email{ozacik@science.ankara.edu.tr}
\affiliation{Department of Physics, Ankara University,\\
 Faculty of Sciences, 06100, Tandogan-Ankara, Turkey}

\author{\"{U}mit Ertem}
\email{umitertem@ankara.edu.tr}
\affiliation{Department of Software Engineering, Ankara University,\\
Faculty of Engineering, 06830 G\"olba\c{s}{\i}, Ankara, Turkey}

\author{\"{O}zg\"{u}r Kelek\c{c}i}
\email{okelekci@thk.edu.tr}
\affiliation{Department of Basic Sciences, Faculty of Engineering,\\
 University of Turkish Aeronautical Association, Ankara, Turkey}

\begin{abstract}

Spinor bilinears of generalized geometry spinors and their properties are investigated. Generalized Killing and twistor spinor equations are considered and their relations to the equations satisfied by special types of differential forms are found. Killing equation in generalized geometry is written in terms of the generalized covariant derivative and Killing-Yano forms are described in the framework of generalized geometry. Construction of generalized Killing-Yano forms and generalized closed conformal Killing-Yano forms in terms of the spinor bilinears of generalized Killing spinors are determined.

\end{abstract}

\maketitle

\section{Introduction}

Generalized geometry is first defined as the extension of the tangent bundle of a manifold to the direct sum of the tangent and cotangent bundles \cite{Hitchin, Gualtieri}. In this way, besides the metric, higher-degree forms can also be incorporated into the geometrical picture. The importance of this approach is that it can be directly used in the geometrical formulation of  supergravity theories in various dimensions \cite{CoimbraConstableWaldram, CoimbraConstableWaldram2}. By the definition of the generalized tangent bundle, the symmetry transformations of the structure group naturally include the closed 2-forms (B-field) \cite{Gualtieri, CoimbraConstableWaldram}. Therefore, the fluxes in supergravity theories, especially in 10-dimensional IIA and IIB theories, can be naturally described as part of the geometry by approaching them from the perspective of generalized geometry. Namely, generalized geometry gives way to describe the supergravity theories in a full geometrical framework. Besides the generalized vectors and generalized metric, one can also define generalizations of the Lie bracket in generalized geometry framework which give the Courant algebroid structure to the generalized tangent bundle \cite{Courant, Roytenberg}. The geometrical structure of supergravity theories can be analyzed more extensively by exploiting the definition of generalized G-structures and generalized spinors  \cite{GranaMinasianPetriniTomasiello, CoimbraConstableWaldram3, JeschekWitt, Witt}. The approach of generalized geometry can also be extended to exceptional cases by defining the generalized tangent bundle to include higher-degree form spaces \cite{Hull, PachecoWaldram, GranaOrsi, AshmorePetriniWaldram}.

Generalized Lie derivative in generalized geometry can be described as the Dorfman bracket of generalized vectors \cite{CoimbraConstableWaldram}. Generalized vectors that preserve the generalized metric with respect to the generalized Lie derivative are called generalized Killing vectors which are the generalized geometry versions of Killing vector fields corresponding to isometries of the ambient manifold \cite{CoimbraConstable}. Construction of generalized geometry spinors and analysis of Killing superalgebra structures in generalized geometry can be realized by using generalized Killing vectors. In that way, the structure of supergravity backgrounds can be described by using generalized Killing superalgebra structures \cite{CoimbraConstable}. Killing vector fields have antisymmetric generalizations called Killing-Yano (KY) forms \cite{Yano, TachibanaKashiwada} and spinor bilinears constructed out of KY forms can allow extending Killing superalgebra structures to include higher-degree KY forms \cite{Ertem, Ertem4, Ertem2, AcikErtem}. These extended Killing superalgebra structures have potential to provide the refinements of the analysis of the structure of supergravity backgrounds. However, description of KY forms and construction of spinor bilinears in terms of KY forms in the context of generalized geometry have been unexplored so far.

In this paper, we construct generalized KY forms in generalized geometry and analyze their relations with generalized geometry Killing spinors. Properties of generalized spinor bilinears are determined for various cases and special spinor equations in generalized geometry are investigated. Special types of spinors such as geometric Killing spinors and twistor spinors which are related to supersymmetry generators of different supersymmetric theories are extended to the framework of generalized geometry via generalized connection and the equations satisfied by them are analyzed in terms of covariant derivative on the exterior bundle. Starting from the description of generalized Killing vectors in terms of generalized Lie derivative, their description in terms of generalized connection is obtained and generalized KY forms are defined by using this description. Relations between generalized geometry Killing spinors and generalized KY forms are investigated and it is found that generalized KY forms and generalized closed conformal KY forms can be constructed from the bilinear forms of generalized geometry Killing spinors. It is shown that bilinear forms of generalized Killing spinors correspond to conformal KY forms in general.

The paper is organized as follows. In Section 2, we summarize the basic formulas and definitions in generalized geometry which are used throughout the paper. Section 3 deals with the choices of spinor inner products for generalized spinors and the properties of generalized spinor bilinears constructed out of them. In Section 4, we consider generalized Killing and twistor spinors and find their description in terms of differential forms of the exterior bundle on the ambient manifold. Section 5 includes the description of generalized Killing vectors in terms of generalized connection and definition of generalized KY forms. In Section 6, relation between generalized Killing spinors and generalized KY forms is investigated. Section 7 concludes the paper.

\section{Generalized Geometry}

Let us consider an $n$-dimensional manifold $M$. We denote the tangent and cotangent bundles whose sections correspond to vector fields and 1-forms as $TM$ and $T^*M$, respectively. In general, $p$-form bundle on $M$ is denoted by $\Lambda^p M$ and $p$-vector bundle is denoted by $\Lambda_p M$. The basic idea of generalized geometry is to combine vectors and 1-forms into a single entity. Therefore, we consider the direct sum of tangent and cotangent bundles $E=TM\oplus T^*M$ which is called the generalized  tangent bundle. Sections of $E$ which is denoted by $\Gamma E$ are called generalized vectors and a generalized vector $\mathcal{X}\in\Gamma E$ can be written in terms of a vector field $X\in \Gamma TM$ and a 1-form $\xi\in\Gamma T^*M$ as
\[
\mathcal{X}=X+\xi
\]
or we can also denote it in the matrix form as $\mathcal{X}=\left(\begin{array}{cc}X\\ \xi\end{array}\right)$. One can define a bilinear form $<\,,\,>$ on $E$ which is written for generalized vectors $\mathcal{X}=X+\xi$ and $\mathcal{Y}=Y+\eta$ as follows
\begin{equation}\label{bilinear1}
<\mathcal{X}, \mathcal{Y}>=\frac{1}{2}\big(i_X\eta+i_Y\xi\big)
\end{equation}
where $i_X$ denotes the interior derivative or contraction with respect to the vector field $X$. So, for a single generalized vector, we have $<\mathcal{X}, \mathcal{X}>=i_X\xi$. This bilinear form has signature $(n, n)$ and with the orientation defined for $E$, the structure group of $E$ reduces to $SO(n, n)$ \cite{Gualtieri}. The invariance of $<\,,\,>$ under $SO(n,n)$ gives shear transformations for generalized vectors. For a 2-form $B\in\Lambda^2M$ and a 2-vector $\beta\in\Lambda_2 M$, we have $B$- and $\beta$-transforms of $\mathcal{X}$ as follows \cite{Gualtieri}
\begin{eqnarray}\label{eq2}
e^B(X+\xi)&=&X+\xi+i_XB\\
e^{\beta}(X+\xi)&=&X+i_{\xi}\beta+\xi
\end{eqnarray}
where $i_{\xi}$ denotes the contraction of a $p$-vector with respect to the 1-form $\xi$.

Note that $e^B$ and $e^\beta$ transformations preserve both orientation and inner product given in \eqref{bilinear1}. A metric $g$ defined on $M$ can be seen as a map $g:TM\rightarrow T^*M$ which is invertible. So, we can define a generalized metric on $E$ induced by $g$ as follows
\begin{equation}\label{eq4}
\mathcal{G}=\left(\begin{array}{cc}0 & g^{-1}\\ g & 0\end{array}\right).
\end{equation}
So, the $\mathcal{G}$-dual of a generalized vector $\mathcal{X}=X+\xi$ can be written as
\[
\widetilde{\mathcal{X}}:=\mathcal{G}(\mathcal{X})=\left(\begin{array}{cc}g^{-1}(\xi)\\ g(X)\end{array}\right)=\left(\begin{array}{cc}\widetilde{\xi}\\ \widetilde{X}\end{array}\right)\in \Gamma E
\]
where $\widetilde{X}$ denotes the $g$-dual 1-form and $\widetilde{\xi}$ denotes the $g$-dual vector field and by the isomorphism $E^*\simeq E$, we have $\Gamma E\simeq \Gamma E^*$ and we can also write $\widetilde{\mathcal{X}}=\left(\begin{array}{cc}\widetilde{X}\\ \widetilde{\xi}\end{array}\right)\in \Gamma E^*$. In general, the $B$-transform of $\mathcal{G}$ can be written as
\begin{equation}\label{eq5}
\mathcal{G}_B=\left(\begin{array}{cc}-g^{-1}B & g^{-1}\\ g-Bg^{-1}B & Bg^{-1}\end{array}\right)
\end{equation}
which is the generalized metric induced by $g$ and $B$. Since $\mathcal{G}^2=I$, $\pm 1$ eigenspaces of $\mathcal{G}$ which are denoted by $V_{\pm}$ give a metric splitting of $E$
\[
E=TM\oplus T^*M=V_+\oplus V_-
\]
corresponding to maximally positive/negative definite subbundles. The generalized metric $\mathcal{G}$ can be written in terms of the bilinear form restricted to $V_{\pm}$ as follows
\begin{equation}\label{eq6}
\mathcal{G}(\,,\,)=<\,,\,>_+-<\,,\,>_-.
\end{equation}
We can also define an admissible metric for which the choices of $V_{\pm}$ correspond to
\[
V_{\pm}=\left\{X\pm\widetilde{X}:X\in\Gamma TM\right\}.
\]

From the basis vectors $\{X_a\}\in\Gamma TM$ and basis 1-forms $\{e^a\}\in\Gamma T^*M$ for $a=1,...,n$, we can define a basis for generalized vectors as
\[
\mathcal{X}_A=X_a+\epsilon e^a
\]
where $\epsilon=\pm 1$ for $V_{\pm}$ and $A=1,...,2n$. If the generalized vector basis $\mathcal{X}_A\in V_+$ then $\epsilon=+1$ and if $\mathcal{X}_A\in V_-$ then $\epsilon=-1$. Since $A$ takes values of 1,...,2n and $a$ takes values of 1,...,n, the above equation is equivalent to
\[
\mathcal{X}_A=\{X_a+e^a,X_{n+a}-e^{n+a}\}
\]
and the left hand side and right hand side indices are equivalent in number and the index $A=\{a, n+a\}$ is just a nomenclature of the right hand side. $\mathcal{G}$-duals of $\mathcal{X}_A$ correspond to basis for generalized 1-forms
\[
\mathcal{E}^A=\mathcal{G}(\mathcal{X}_A)=e^a+\epsilon X_a.
\]
Then, we can define the elements of $\Lambda^p E^*$ as generalized $p$-forms and a generalized $p$-form $\mathcal{A}$ can be written in terms of the generalized 1-form basis as
\[
\mathcal{A}=a_{A_1A_2...A_p}\mathcal{E}^{A_1}\wedge\mathcal{E}^{A_2}\wedge ... \wedge \mathcal{E}^{A_p}
\]
where $a_{A_1A_2...A_p}$ is a function on $M$.

Similar to the Lie bracket $[\,,\,]$ of vector fields on $\Gamma TM$, we can also define a bracket operation on $E$ called Courant bracket as $[\,,\,]_C:\Gamma E\times\Gamma E\rightarrow \Gamma E$. For two generalized vectors $\mathcal{X}=X+\xi$ and $\mathcal{Y}=Y+\eta$, Courant bracket is defined as
\begin{equation}\label{eq7}
[\mathcal{X}, \mathcal{Y}]_C=[X,Y]+\mathcal{L}_X\eta-\mathcal{L}_Y\xi-\frac{1}{2}d\left(i_X\eta-i_Y\xi\right)
\end{equation}
where $\mathcal{L}_X$ is the Lie derivative with respect to the vector field $X$ and $d$ is the exterior derivative. If we define the anchor map $\pi:\Gamma E\rightarrow\Gamma TM$ as $\pi(\mathcal{X})=X$, then the Courant bracket satisfies
\begin{equation}\label{eq8}
\pi([\mathcal{X},\mathcal{Y}]_C)=[\pi(\mathcal{X}), \pi(\mathcal{Y})]
\end{equation}
which can easily be seen from the definition. Although the Courant bracket is an antisymmetric bracket, it does not satisfy the Jacobi identity and hence does not correspond to a Lie bracket. However, with the definition of the Courant bracket, $E$ admits a Courant algebroid structure \cite{Roytenberg}. One can define an isotropic splitting on $E$, such that $s:\Gamma TM\rightarrow \Gamma E$ and this determines a closed 3-form $H$ on $M$ given by
\begin{equation}\label{eq9}
H(X,Y,Z)=2<[s(X),s(Y)]_C,s(Z)>
\end{equation}
for $X,Y,Z\in \Gamma TM$. The cohomology class of $H$ classifies the Courant algebroids on $M$ up to isomorphism \cite{BresslerChervov, SeveraWeinstein, Bouwknegt}. In the presence of non-zero $H$, one can modify the Courant bracket and define the twisted Courant bracket $[\,,\,]_H$ as follows
\begin{equation}\label{eq10}
[\mathcal{X},\mathcal{Y}]_H=[\mathcal{X},\mathcal{Y}]_C-i_Xi_YH.
\end{equation}
$[\,,\,]_H$ also defines a Courant algebroid structure for $dH=0$.
Another bracket operation on $E$ which satisfies the Jacobi identity but is not antisymmetric is the Dorfman bracket $[\,,\,]_D$ and it is defined as
\begin{equation}\label{eq11}
[\mathcal{X}, \mathcal{Y}]_D=[X,Y]+\mathcal{L}_X\eta-i_Yd\xi.
\end{equation}
Courant bracket corresponds to the antisymmetrization of the Dorfman bracket and it can also be written as
\begin{equation}\label{eq12}
[\mathcal{X}, \mathcal{Y}]_C=[\mathcal{X}, \mathcal{Y}]_D-d<\mathcal{X}, \mathcal{Y}>.
\end{equation}
Dorfman bracket originates from the Lie derivative on differential forms and it is also considered as the generalized Lie derivative
\[
\mathbb{L}_{\mathcal{X}}\mathcal{Y}=[\mathcal{X}, \mathcal{Y}]_D
\]
and it satisfies \cite{Baraglia}
\[
\mathbb{L}_{\mathcal{X}}({f\mathcal{Y}})=(\pi(\mathcal{X})f)\mathcal{Y}+f\mathbb{L}_{\mathcal{X}}\mathcal{Y}
\]
\begin{equation}\label{eq13}
[\mathbb{L}_{\mathcal{X}}, \mathbb{L}_{\mathcal{Y}}]=\mathbb{L}_{[\mathcal{X},\mathcal{Y}]_D}
\end{equation}
\[
\mathbb{L}_{\mathcal{X}}(\mathcal{Y}\otimes \mathcal{Z})=\mathbb{L}_{\mathcal{X}}\mathcal{Y}\otimes\mathcal{Z}+\mathcal{Y}\otimes\mathbb{L}_{\mathcal{X}}\mathcal{Z}.
\]
where $f$ is a function and $\mathcal{X}, \mathcal{Y}, \mathcal{Z} \in \Gamma E$.

Generalized tangent bundle $E$ has a natural action on the exterior bundle $\Lambda M$ of $M$. For $\mathcal{X}=X+\xi\in\Gamma E$ and $\omega\in\Lambda M$, it is given by
\begin{equation}\label{eq14}
(X+\xi).\omega=i_X\omega+\xi\wedge\omega.
\end{equation}
Then, it can easily be seen that
\[
(X+\xi)^2.\omega=<\mathcal{X},\mathcal{X}>\omega
\]
which defines a Clifford algebra structure on $E$. So, the Clifford algebra has a natural representation on $S=\Lambda M$ which is the generalized spinor space. Since for signature $(n,n)$, the volume element $z$ satisfies $z^2=1$, we have
\[
S=S^+\oplus S^-
\]
which is equivalent to the decomposition
\[
\Lambda M=\Lambda^{even}M\oplus\Lambda^{odd}M.
\]
Hence, the positive helicity generalized spinors on $E$ correspond to even degree differential forms on $M$ and negative helicity generalized spinors on $E$ correspond to odd degree differential forms on $M$. One can also define a spin invariant inner product on $S$ which is discussed in Section 3.

We can define a generalized connection $\mathbb{D}$ on $E$ which is compatible with the bilinear form $<\,,\,>$ corresponding to the linear operator
\[
\mathbb{D}:\Gamma E\rightarrow\Gamma(E^*\otimes E)
\]
satisfying the following identities
\begin{eqnarray}\label{eq15}
\mathbb{D}_{\mathcal{X}}(f\mathcal{Y})&=&\pi(\mathcal{X})(f)\mathcal{Y}+f\mathbb{D}_{\mathcal{X}}\mathcal{Y}\\
\pi(\mathcal{X})<\mathcal{Y},\mathcal{Z}>&=&<\mathbb{D}_{\mathcal{X}}\mathcal{Y},\mathcal{Z}>+<\mathcal{Y},\mathbb{D}_{\mathcal{X}}\mathcal{Z}>
\end{eqnarray}
where $f$ is a function, $\mathcal{X},\mathcal{Y},\mathcal{Z}\in\Gamma E$ and we denote $\mathbb{D}_{\mathcal{X}}\mathcal{Y}=i_{\mathcal{X}}\mathbb{D}(\mathcal{Y})$. Here $i$ is the contraction operator of elements in $\Gamma E^*$ with elements in $\Gamma E$. Given a standard connection $\nabla$ on $TM$, we can define the generalized connection $\mathbb{D}^{\nabla}$ induced by $\nabla$ as
\[
\mathbb{D}^{\nabla}=\nabla\oplus\nabla
\]
corresponding to
\begin{equation} \label{eq17}
\mathbb{D}^{\nabla}_{\mathcal{X}}\mathcal{Y}=\nabla_{\pi(\mathcal{X})}\mathcal{Y}.
\end{equation}
However, there is a freedom in choosing a connection since $\mathbb{D}'=\mathbb{D}+\alpha$ also defines a connection where $\alpha\in\Gamma (E^*\otimes o(E))$ and $o(E)$ denotes the bundle of skew-symmetric endomorphisms of $E$ w.r.t. $<\,,\,>$ \cite{GarciaFernandez, Gualtieri2}. The generalized torsion 3-form $T_{\mathbb{D}}\in\Gamma (\Lambda^3E^*)$ of a generalized connection $\mathbb{D}$ on $E$ is defined by
\[
T_{\mathbb{D}}(\mathcal{X},\mathcal{Y},\mathcal{Z})=<\mathbb{D}_{\mathcal{X}}\mathcal{Y}-\mathbb{D}_{\mathcal{Y}}\mathcal{X}-[\mathcal{X},\mathcal{Y}]_D,\mathcal{Z}>+<\mathbb{D}_{\mathcal{Z}}\mathcal{X},\mathcal{Y}>
\]
or in terms of the Courant bracket
\begin{equation}\label{eq18}
T_{\mathbb{D}}(\mathcal{X},\mathcal{Y},\mathcal{Z})=<\mathbb{D}_{\mathcal{X}}\mathcal{Y}-\mathbb{D}_{\mathcal{Y}}\mathcal{X}-[\mathcal{X},\mathcal{Y}]_C,\mathcal{Z}>+\frac{1}{2}\left(<\mathbb{D}_{\mathcal{Z}}\mathcal{X},\mathcal{Y}>-<\mathbb{D}_{\mathcal{Z}}\mathcal{Y},\mathcal{X}>\right).
\end{equation}

In the presence of a non-zero 3-form field $H$, the conditions of metric compatibility and vanishing generalized torsion do not uniquely fix the connection; rather, they define a class of admissible connections. However, as discussed in \cite{CoimbraConstable} (see Eq. 3.4 therein), the relevant physical operators acting on spinors are defined intrinsically by the generalized exterior derivative. Therefore, the resulting equations are independent of the specific choice of connection within this class. Consequently, for explicit calculations, we are free to adopt the canonical choice of the generalized connection \cite{GarciaFernandez} without loss of generality:

\begin{equation}\label{eq19}
\mathbb{D}^{H}_{\mathcal{X}}\mathcal{Y}=\nabla_{\pi(\mathcal{X})}\mathcal{Y}+\epsilon\frac{1}{6}i_{\mathcal{Y}}i_{\pi(\mathcal{X})}H
\end{equation}
where $\mathbb{D}^{H}$ acts on the subbundles $V_{\pm}$ for $\epsilon=\pm 1$, respectively. For $\mathcal{X}=X+\xi$, $\mathcal{Y}=Y+\eta\in\Gamma E$, it can be written explicitly as
\begin{equation}\label{eq20}
\mathbb{D}^{H}_{\mathcal{X}}\mathcal{Y}=\nabla_X\left(\begin{array}{cc}Y\\ \eta\end{array}\right)+\epsilon\frac{1}{6}\left(\begin{array}{cc}\widetilde{i_Yi_X H}\\ i_{\widetilde{\eta}}i_X H\end{array}\right).
\end{equation}

By using the generalized connection, generalized 1-form basis forms $\{\mathcal{E}^A\}$ and the Clifford product, one can also define the generalized Dirac operator acting on generalized spinors $\phi\in S$ as follows
\begin{equation}\label{eq21}
\displaystyle{\not}\mathbb{D}^H\phi=\mathcal{E}^A.\mathbb{D}^H_{\mathcal{X}_A}\phi
\end{equation}
where the induced connection on generalized spinors is also denoted by $\mathbb{D}^H$.

\section{Inner product classes and Properties of Generalized Spinor Bilinears}

For the generalized tangent bundle $E$, we have signature $(n,n)$ and the Clifford algebra $Cl(n,n)$ is isomorphic to
\begin{equation}\label{eq22}
Cl(n,n)\simeq \mathbb{R}(2^n)
\end{equation}
where $\mathbb{R}(2^n)$ denotes the real matrices with $2^n$ entries. The even subalgebra $Cl^0(n,n)$ of $Cl(n,n)$ which is isomorphic to the Clifford algebra in one lower dimension $Cl^0(n,n)\simeq Cl(n,n-1)$ corresponds to
\begin{equation}\label{eq23}
Cl^0(n,n)\simeq \mathbb{R}(2^{n-1})\oplus\mathbb{R}(2^{n-1}).
\end{equation}
So, the corresponding spinor space is isomorphic to
\begin{equation}\label{eq24}
S\simeq \mathbb{R}^{2^{n-1}}\oplus\mathbb{R}^{2^{n-1}}
\end{equation}
and the spinors are Majorana-Weyl spinors.
In our case, we have
\begin{equation}\label{eq25}
S=S^+\oplus S^-=\Lambda^{even}M\oplus\Lambda^{odd}M.
\end{equation}
which is isomorphic to $\mathbb{R}^{2^{n-1}}\oplus\mathbb{R}^{2^{n-1}}$. The spinor inner product $(\,,\,):S\times S\rightarrow \mathbb{R}$ of two generalized geometry spinors $\phi$ and $\psi$ is written as
\begin{eqnarray}\label{eq26}
(\phi, \psi) &=& (J^{-1}\phi^{\mathcal{J}}.\psi)_0 z\nonumber\\
&=&(\phi^{\mathcal{J}}\wedge \psi)_n
\end{eqnarray}
where $J$ is any matrix that we choose as identity matrix, $\mathcal{J}$ is an inner automorphism of the Clifford algebra and $(\,\,)_n$ denotes the projection to the $n$-form part. For $\mathcal{J}$, we have two choices which are $\xi$ and $\xi\eta$ automorphisms where $\eta$ acts on a $p$-form $\omega$ as $\omega^{\eta}=(-1)^p\omega$ and $\xi$ acts as $\omega^{\xi}=(-1)^{\lfloor \frac{p}{2}\rfloor}\omega$ with $\xi\eta$ is the composition of both automorphisms. Here, $\lfloor \rfloor$ denotes the floor function. We can determine the symmetry properties and automorphism classes of inner products for different values of $n$ by using the spinor inner product classification tables given in \cite{BennTucker, ErtemSutemenAcikCatalkaya} as follows

\quad \quad \quad \quad Table-1. Spinor inner product classification table 
\\
{\centering{ 
\begin{tabular}{c c c c}

% after \\: \hline or \cline{col1-col2-col3}...
\hline 
$n(\text{mod }4)$ & $\text{ }$ & $\text{ }$\\ \hline
$0$ & $\mathbb{R}\text{-swap  } \xi$ & $\text{  and  }$ & $\mathbb{R}\text{-sym}\oplus\mathbb{R}\text{-sym  }\xi\eta$ \\
$1$ & $\mathbb{R}\text{-sym}\oplus\mathbb{R}\text{-sym  }\xi$ & $\text{  and  }$ & $\mathbb{R}\text{-swap  } \xi\eta$ \\
$2$ & $\mathbb{R}\text{-swap  } \xi$ & $\text{  and  }$ & $\mathbb{R}\text{-skew}\oplus\mathbb{R}\text{-skew  }\xi\eta$\\
$3$ & $\mathbb{R}\text{-skew}\oplus\mathbb{R}\text{-skew  }\xi$ & $\text{  and  }$ & $\mathbb{R}\text{-swap  } \xi\eta$\\ \hline 

\end{tabular}
\quad\\
\quad\\}}

Since the spinor space is real, we have real inner products, sym and skew denotes the symmetric and skew inner products and swap means that inner product changes the chiral elements to each other, $\xi$ and $\xi\eta$ correspond to the relevant automorphism class. So, we have for $n=0,1 (\text{mod  } 4)$ symmetric and for $n=2,3 (\text{mod  } 4)$ skew inner products. Since, only the $n$-form part remains in the spinor inner product, one can see that, for $n=0(\text{mod  }2)$, inner products of elements
\[
S^+ \times S^- \text{  and  } S^-\times S^+ \text{  are zero  }
\]
\[
S^+ \times S^+ \text{  and  } S^-\times S^- \text{  are non-zero  } 
\]
since the inner products of opposite chirality spinors cannot include even degree part and for $n=1(\text{mod  }2)$, inner products of elements
\[
S^+ \times S^+ \text{  and  } S^-\times S^- \text{  are zero  }
\]
\[
S^+ \times S^- \text{  and  } S^-\times S^+ \text{  are non-zero  } 
\]
since the inner products of same chirality spinors cannot include odd degree part. Then, for $n=0(\text{mod  }2)$, we need to choose $\mathbb{R}\text{-sym}\oplus\mathbb{R}\text{-sym}$ or $\mathbb{R}\text{-skew}\oplus\mathbb{R}\text{-skew}$ inner products since only the inner products of same chirality spinors are non-zero in this case and for $n=1(\text{mod  }2)$, we need to choose $\mathbb{R}\text{-swap}$ inner products since only the inner products of opposite chirality spinors are non-zero in this case. As a result, we have the inner product symmetries and automorphisms classes as follows

\quad\\
{\centering{ Table-2. Spinor inner product symmetries and automorphism classes \\
\begin{tabular}{c c}

% after \\: \hline or \cline{col1-col2-col3}...
\hline 
$n(\text{mod }4)$ & $\text{ }$\\ \hline
$0$ & $\mathbb{R}\text{-sym}\oplus\mathbb{R}\text{-sym  }\xi\eta$ \\
$1$ & $\mathbb{R}\text{-swap  (sym)  } \xi\eta$ \\
$2$ & $\mathbb{R}\text{-skew}\oplus\mathbb{R}\text{-skew  }\xi\eta$\\
$3$ & $\mathbb{R}\text{-swap  (skew)  } \xi\eta$\\ \hline 

\end{tabular}
\quad\\
\quad\\
\quad\\}}

Hence, we can write the spinor inner product for generalized spinors as
\begin{equation}\label{eq27}
(\phi,\psi)=(\phi^{\xi\eta}\wedge\psi)_n
\end{equation}
which corresponds to the Mukai pairing of forms. It can be seen from the previous analysis of inner product classes that the automorphism class must be taken as $\xi\eta$ to have a non-zero spinor inner product. For any two generalized spinors $\phi$, $\psi$ and a generalized form $\mathcal{\alpha}$, we can write
\begin{equation}\label{eq28}
(\phi, \alpha.\psi)=(\mathcal{\alpha}^{\mathcal{J}}.\phi, \psi)
\end{equation}
where the action of a generalized form on a generalized geometry spinor $\alpha.\psi$ can be defined inductively  from the action of a generalized vector on a differential form given in \eqref{eq14} . From the symmetry properties of the inner product and the choice of the automorphism class as $\xi\eta$, we have
\begin{equation}\label{eq29}
(\phi, \alpha.\psi)=(-1)^{n(n-1)/2}(\psi, \mathcal{\alpha}^{\xi\eta}.\phi).
\end{equation}
For a spinor $\phi$ and a $p$-form $\alpha_p$, we can define the spinor bilinear $p$-form as $(\phi, \alpha_p.\phi)$ and writing the spinor bilinears for a generalized geometry spinor $\phi$ and a generalized $p$-form $\mathcal{\alpha}_p$, we have
\begin{eqnarray}\label{eq30}
(\phi, \mathcal{\alpha}_p.\phi)&=&(-1)^{n(n-1)/2}(\phi, \mathcal{\alpha}^{\xi\eta}_p.\phi)\nonumber\\
&=&(-1)^{n(n-1)/2}(-1)^{p(p-1)/2}(-1)^p(\phi, \mathcal{\alpha}_p.\phi)\nonumber\\
&=&(-1)^{\frac{n(n-1)+p(p+1)}{2}}(\phi, \mathcal{\alpha}_p.\phi)
\end{eqnarray}
where we have used $\mathcal{\alpha}_p^{\xi\eta}=(-1)^{p(p-1)/2}(-1)^p\mathcal{\alpha}_p$. Moreover, for $n$ even, $p$ must be even, and for $n$ odd, $p$ must be odd, since for $n$ even $S^\pm \times S^\pm$ and for $n$ odd $S^\pm \times S^\mp$ inner products are non-zero. For a definite chirality spinor $\phi$, since the action of a generalized vector on a generalized geometry spinor changes the chirality of the spinor, $\alpha_p.\phi$ has the same chirality if $p$ is even and has the reverse chirality if $p$ is odd. Then, by considering those properties, $(-1)^{\frac{n(n-1)+p(p+1)}{2}}=1$ only for $n+p=0  (\text{mod  }4)$ and
\[
(\phi, \mathcal{\alpha}_p.\phi)
\]
is non-zero and for other cases the bilinear form is zero. In general, for two different generalized geometry spinors $\phi$ and $\psi$ $(\phi\neq \psi)$, we have
\[
(\phi, \mathcal{\alpha}_p.\psi)=(\psi, \mathcal{\alpha}_p.\phi)\text{\,\,\,\,\,for  }n+p=0(\text{mod  }4)
\]
\[
(\phi, \mathcal{\alpha}_p.\psi)=-(\psi, \mathcal{\alpha}_p.\phi)\text{\,\,\,\,\,for  }n+p=2(\text{mod  }4)
\]
and other cases are zero. Then, only some special degree spinor bilinears are non-zero depending on $n$ and $p$. For two generalized geometry spinors $\phi$ and $\psi$, we can write the spinor bilinear as the tensor product of a dual spinor and a spinor which can be written as a sum of $p$-form bilinears in terms of generalized $p$-form basis as
\begin{equation}\label{eq31}
\overline{\psi}\otimes\phi=(\phi,\psi)+(\phi, \mathcal{E}_A.\psi)\mathcal{E}^A+...+(\phi, \mathcal{E}_{A_p...A_2A_1}.\psi)\mathcal{E}^{A_1A_2...A_p}+...
\end{equation}
where $\overline{\psi}$ denotes the dual spinor and the action of generalized connection $\mathbb {D}$ on the bilinear $p$-form can be written as
\begin{equation}\label{eq32}
\mathbb{D}_{\mathcal{X}}(\phi,\mathcal{E}_{A_p...A_2A_1}.\psi)=(\mathbb{D}_{\mathcal{X}}\phi,\mathcal{E}_{A_p...A_2A_1}.\psi)+(\phi, (\mathbb{D}_{\mathcal{X}}\mathcal{E}_{A_p...A_2A_1}).\psi)+(\phi, \mathcal{E}_{A_p...A_2A_1}.\mathbb{D}_{\mathcal{X}}\psi).
\end{equation}

\section{Generalized Killing and twistor spinors}

We now consider the action of the generalized connection on spinors. It is worth noting that while the supersymmetry parameters in supergravity backgrounds are typically sections of the spinor bundles associated with the $V_{\pm}$ splitting of the generalized metric, we here investigate the properties of the full $SO(n,n)$ spinors. These are representable as inhomogeneous differential forms on $M$. We treat these objects as the natural mathematical generalization of geometric spinors within the framework of the generalized tangent bundle $E$.

The action of the generalized connection $\mathbb{D}^H$ with respect to the generalized vector basis $\mathcal{X}_A = X_a + \epsilon e^a$ on a generalized vector $\mathcal{Y} = Y + \eta$ is written as

\begin{equation}\label{eq33}
\mathbb{D}^H_{\mathcal{X}_A}\mathcal{Y}=\nabla_{X_a}\left(\begin{array}{cc}Y\\ \eta\end{array}\right)+\epsilon\frac{1}{6}\left(\begin{array}{cc}\widetilde{i_Yi_{X_a}H}\\ i_{\widetilde{\eta}}i_{X_a}H\end{array}\right)
\end{equation}
where $\epsilon=\pm$ on subbundles $V_{\pm}$. In fact, the second term on the right hand side corresponds to the Clifford bracket action of $-\frac{1}{2}i_{X_a}H$ to 1-forms $\widetilde{Y}$ and $\eta$
\begin{equation}\label{eq34}
\mathbb{D}^H_{\mathcal{X}_A}\mathcal{Y}=\nabla_{X_a}\mathcal{Y}-\epsilon\frac{1}{12}[i_{X_a}H, \mathcal{Y}]_{Cl}
\end{equation}
where we can write the Clifford bracket of a 2-form $\alpha$ with any form $\omega$ as \cite{BennTucker}
\begin{equation}\label{eq35}
[\alpha,\omega]_{Cl}=-2i_{X_a}\alpha\wedge i_{X^a}\omega
\end{equation}
and for the cases of 2-form $-\frac{1}{2}i_{X_a}H$ and 1-forms $\widetilde{Y}$ and $\eta$, we have
\begin{eqnarray}\label{eq36}
-\frac{1}{2}[i_{X_a}H, \widetilde{Y}]_{Cl}&=&i_{X_b}i_{X_a}H\wedge i_{X^b}\widetilde{Y}=i_Yi_{X_a}H\\
-\frac{1}{2}[i_{X_a}H, \eta]_{Cl}&=&i_{X_b}i_{X_a}H\wedge i_{X^b}\eta=i_{\widetilde{\eta}}i_{X_a}H.
\end{eqnarray}
They correspond to the second term on the right hand side of \eqref{eq33}. and we define the Clifford bracket on the right hand side of \eqref{eq34} as
\begin{equation}\label{eqn1}
[i_{X_a}H, \mathcal{Y}]_{Cl}=\left(\begin{array}{cc}\widetilde{[i_{X_a}H, \widetilde{Y}]}_{Cl}\\{[i_{X_a}H, {\eta}]}_{Cl} \end{array}\right).
\end{equation}

Let us consider a generalized geometry spinor $\psi$ corresponding to an inhomogeneous differential form which satisfies the Killing spinor equation in terms of the generalized connection and generalized 1-forms
\begin{equation}\label{eq38}
\mathbb{D}^H_{\mathcal{X}_A}\psi=\lambda \mathcal{E}_A.\psi
\end{equation}
where $\lambda$ is a real or pure imaginary number called the Killing number. This equation corresponds to the generalization of the geometric Killing spinor equation
\[
\nabla_{X_a}\psi=\lambda e_a.\psi
\]
to generalized geometry. Then, we can write the left hand side of \eqref{eq38} as
\begin{eqnarray}\label{eq39}
\mathbb{D}^H_{\mathcal{X}_A}\psi&=&\nabla_{X_a}\psi-\epsilon\frac{1}{12}[i_{X_a}H, \psi]_{Cl}\nonumber\\
&=&\nabla_{X_a}\psi+\epsilon\frac{1}{6}i_{X_b}i_{X_a}H\wedge i_{X^b}\psi\nonumber\\
&:=&\nabla^H_{X_a}\psi
\end{eqnarray}
where we have defined the connection $\nabla^H_{X_a}=\nabla_{X_a}+\epsilon\frac{1}{6}i_{X_b}i_{X_a}H\wedge i_{X^b}$. For the right hand side of the Killing spinor equation in generalized geometry, by using \eqref{eq14}, we have
\begin{eqnarray}\label{eq40}
\lambda \mathcal{E}_A.\psi&=&\lambda(e_a+\epsilon X_a).\psi\nonumber\\
&=&\lambda(\epsilon i_{X_a}\psi+e_a\wedge \psi)\nonumber\\
&=&\biggl\{  \begin{array}{rcl} \lambda e_a.\psi & \text{for} & \epsilon=+1\\ \lambda \psi^{\eta}.e_a & \text{for} & \epsilon=-1 \end{array}
\end{eqnarray}
where we have used the Clifford action of 1-form $e_a$ on any differential form $\omega$ as
\begin{eqnarray}\label{eq41}
e_a.\omega&=&e_a\wedge \omega+i_{X_a}\omega\\
\omega.e_a&=&e_a\wedge \omega^{\eta}-i_{X_a}\omega^{\eta}.
\end{eqnarray}
So, the Killing spinor equation in generalized geometry transforms into
\begin{equation}\label{eq43}
\nabla^H_{X_a}\psi=\biggl\{  \begin{array}{rcl} \lambda e_a.\psi & \text{for} & \epsilon=+1\\ \lambda \psi^{\eta}.e_a & \text{for} & \epsilon=-1 \end{array}
\end{equation}
and we can write this in terms of component forms $\psi=\psi_0+\psi_1+...+\psi_p+...$, where $\psi_p$ is a $p$-form, as
\begin{equation}\label{eq44}
\nabla^H_{X_a}(\psi_0+\psi_1+...+\psi_p+...)= \biggl\{  \begin{array}{rcl} \lambda e_a.(\psi_0+\psi_1+...+\psi_p+...) & \text{for} & \epsilon=+1\\ \lambda (\psi_0+\psi_1+...+\psi_p+...)^{\eta}.e_a & \text{for} & \epsilon=-1 \end{array}
\end{equation}
\begin{equation}\label{eq45}
\Rightarrow \nabla^H_{X_a}\psi_p=\lambda(e_a\wedge \psi_{p-1}+\epsilon i_{X_a}\psi_{p+1}).
\end{equation}
If we apply $e^a\wedge$ to \eqref{eq45}, we obtain
\begin{eqnarray}\label{eq46}
e^a\wedge \nabla^H_{X_a}\psi_p&=&\epsilon\lambda e^a\wedge i_{X_a}\psi_{p+1}\nonumber\\
\Rightarrow d^H\psi_p&=&\epsilon \lambda (p+1)\psi_{p+1}\nonumber\\
\Rightarrow i_{X_a}d^H\psi_p&=&\epsilon\lambda(p+1)i_{X_a}\psi_{p+1}
\end{eqnarray}
where we have used the identities $d^H=e^a\wedge \nabla^H_{X_a}$ and $e^a\wedge i_{X_a}\psi_p=p\psi_p$. Similarly, by applying $-i_{X^a}$ to \eqref{eq45}, we obtain
\begin{eqnarray}\label{eq47}
-i_{X^a}\nabla^H_{X_a}\psi_p&=&-\lambda i_{X^a}(e_a\wedge \psi_{p-1})\nonumber\\
\Rightarrow \delta^H\psi_p&=&-\lambda(n-p+1)\psi_{p-1}\nonumber\\
\Rightarrow e^a\wedge \delta^H\psi_p&=&-\lambda(n-p+1)e^a\wedge\psi_{p-1}
\end{eqnarray}
where we have defined $\delta^H=-i_{X^a}\nabla^H_{X_a}$. So, we can write \eqref{eq45} in the following form
\begin{equation}\label{eq48}
\nabla^H_{X_a}\psi_p=\frac{1}{p+1}i_{X_a}d^H\psi_p-\frac{1}{n-p+1}e_a\wedge\delta^H\psi_p
\end{equation}
which corresponds to the conformal Killing-Yano (CKY) equation written in terms of a connection modified by a 3-form flux $H$ corresponding to torsion \cite{HouriKubiznakWarnickYasui,Chrysikos,Ertem3,ChervonyiLunin}. Here, $H$ corresponds to the skew-symmetric torsion and torsion 2-forms $T^a$ are defined as $T^a=\frac{1}{3}i_{X^a}H$. Then, a solution of the generalized geometry Killing spinor equation corresponds to a solution of the CKY equation with torsion $H$.

Similarly, let us consider the generalized geometry twistor spinor equation
\begin{equation}\label{eq49}
\mathbb{D}^H_{\mathcal{X}_A}\psi=\frac{1}{2n}\mathcal{E}_A.\displaystyle{\not}\mathbb{D}\psi
\end{equation}
which is the generalization of the twistor spinor equation
\begin{equation}\label{eq50}
\nabla_{X_a}\psi=\frac{1}{n}e_a.\displaystyle{\not}D\psi
\end{equation}
to generalized geometry. Here $\displaystyle{\not}D$ is the Dirac operator acting on spinors. The left hand side of \eqref{eq49} can be written as
\begin{eqnarray}\label{eq51}
\mathbb{D}^H_{\mathcal{X}_A}\psi&=&\nabla_{X_a}\psi-\epsilon\frac{1}{12}[i_{X_a}H, \psi]_{Cl}\nonumber\\
&=&\nabla^H_{X_a}\psi
\end{eqnarray}
and the generalized Dirac operator acting on generalized geometry spinor $\psi$ corresponds to
\begin{eqnarray}\label{eq52}
\displaystyle{\not}{\mathbb{D}}^H\psi&=&\mathcal{E}^A.{\mathbb{D}}^H_{\mathcal{X}_A}\psi\nonumber\\
&=&(e_a\wedge +\epsilon i_{X_a})\nabla^H_{X^a}\psi\nonumber\\
&=&\{(e_a\wedge +i_{X_a})\nabla^H_{X^a}\psi, (e_{n+a}\wedge - i_{X_{n+a}})\nabla^H_{X^{n+a}}\psi\}\nonumber\\
&=&\{d^H\psi-\delta^H\psi, d^H\psi+\delta^H\psi\}\nonumber\\
&=&d^H\psi-\epsilon\delta^H\psi.
\end{eqnarray}
Thus, the generalized twistor spinor equation can be written as
\begin{eqnarray}\label{eq53}
\nabla^H_{X_a}\psi&=&\frac{1}{2n}\mathcal{E}^A.(d^H\psi-\epsilon\delta^H\psi)\nonumber\\
&=&\frac{1}{2n}\bigg(e^a\wedge(d^H\psi-\epsilon\delta^H\psi)+\epsilon i_{X_a}(d^H\psi-\epsilon\delta^H\psi)\bigg)
\end{eqnarray}
or in terms of components, we have
\begin{equation}\label{eq54}
\nabla^H_{X_a}\psi_p=\frac{1}{2n}\bigg(e_a\wedge d^H\psi_{p-2}+\epsilon i_{X_a}d^H\psi_p-\epsilon e_a\wedge\delta^H\psi_p-i_{X_a}\delta^H\psi_{p+2}\bigg).
\end{equation}
So, a solution of the generalized twistor equation corresponds to a solution of the equation \eqref{eq54}. In particular, if the generalized geometry twistor spinor $\psi$ consists of a homogeneous degree differential form, namely it only includes the $p$-form component $\psi=\psi_p$, then the above equation reduces to
\begin{equation}\label{eqn2}
\nabla^H_{X_a}\psi_p=\frac{1}{2n}\epsilon\bigg(i_{X_a}d^H\psi_p-e_a\wedge\delta^H\psi_p\bigg)
\end{equation}
which resembles the CKY equation with torsion, however the coefficients of the terms on the right hand side are different from the CKY equation.
 
\section{Generalized Killing-Yano equation}

A $p$-form $\omega$ is called a KY $p$-form, if it satisfies the following equation
\begin{equation}\label{eq55}
\nabla_ {X_a}\omega=\frac{1}{p+1}i_{X_a}d\omega
\end{equation}
and they are called hidden symmetries which correspond to the antisymmetric generalizations of Killing vector fields that are isometries of the ambient manifold. We can generalize the concept of KY form to generalized geometry in terms of generalized $p$-forms and generalized connection. With respect to a generalized vector $\mathcal{X}=X+\xi$, the action of generalized Lie derivative and generalized connection on functions are given by

\begin{equation}\label{eq56}
\mathbb{L}_{\mathcal{X}}f=\pi(\mathcal{X})f
\end{equation}
and
\begin{equation}\label{eq57}
\mathbb{D}_{\mathcal{X}}f=\pi(\mathcal{X})f.
\end{equation}
Let us consider the difference operator
\begin{equation}\label{eq58}
\mathcal{A}_{\mathcal{X}}=\mathbb{L}_{\mathcal{X}}-\mathbb{D}_{\mathcal{X}}.
\end{equation}
Then, we have $\mathcal{A}_{\mathcal{X}}f=0$.
By applying $\mathcal{A}_{\mathcal{X}}$ to the generalized metric $\mathcal{G}(\mathcal{Y}, \mathcal{Z})$ defined in \eqref{eq6} for two generalized vectors $\mathcal{Y}, \mathcal{Z}$, we can write
\begin{equation}\label{eq59}
\mathcal{A}_{\mathcal{X}}(\mathcal{G}(\mathcal{Y}, \mathcal{Z}))=0=(\mathcal{A}_{\mathcal{X}}\mathcal{G})(\mathcal{Y}, \mathcal{Z})+\mathcal{G}(\mathcal{A}_{\mathcal{X}}\mathcal{Y}, \mathcal{Z})+\mathcal{G}(\mathcal{Y}, \mathcal{A}_{\mathcal{X}}\mathcal{Z}).
\end{equation}
because $\mathcal{G}(\mathcal{Y}, \mathcal{Z})$ is a function. Since $\mathbb{D}$ is  metric compatible $\mathbb{D}_{\mathcal{X}}\mathcal{G}=0$ and from \eqref{eq58} we have
\begin{equation}\label{eq60}
\mathcal{G}(\mathcal{A}_{\mathcal{X}}\mathcal{Y}, \mathcal{Z})+\mathcal{G}(\mathcal{Y}, \mathcal{A}_{\mathcal{X}}\mathcal{Z})=-(\mathbb{L}_{\mathcal{X}}\mathcal{G})(\mathcal{Y}, \mathcal{Z}).
\end{equation}
We can write the generalized Lie derivative in terms of Dorfman and Courant brackets from \eqref{eq12} as
\begin{eqnarray}\label{eq61}
\mathbb{L}_{\mathcal{X}}\mathcal{Y}&=&[\mathcal{X}, \mathcal{Y}]_D\\
&=&[\mathcal{X}, \mathcal{Y}]_C+d<\mathcal{X}, \mathcal{Y}>.
\end{eqnarray}
So, we can write \eqref{eq58} as
\begin{equation}\label{eq63}
\mathcal{A}_{\mathcal{X}}\mathcal{Y}=[\mathcal{X}, \mathcal{Y}]_C+d<\mathcal{X}, \mathcal{Y} >-\mathbb{D}_{\mathcal{X}}\mathcal{Y}.
\end{equation}
By considering the torsion-free connection $\mathbb{D}^H$, we have from \eqref{eq18}
\begin{equation}\label{eq64}
T_{\mathbb{D}^H}(\mathcal{X}, \mathcal{Y}, \mathcal{Z})=<\mathbb{D}^H_{\mathcal{X}}\mathcal{Y}-\mathbb{D}^H_{\mathcal{Y}}\mathcal{X}-[\mathcal{X}, \mathcal{Y}]_C, \mathcal{Z}>+\frac{1}{2}\left(<\mathbb{D}^H_{\mathcal{Z}}\mathcal{X}, \mathcal{Y}>-<\mathbb{D}^H_\mathcal{Z}\mathcal{Y}, \mathcal{X}>\right)=0.
\end{equation}
One can write the generalized vectors $\mathcal{X}=\mathcal{X}_++\mathcal{X}_-$ and $\mathcal{Y}=\mathcal{Y}_++\mathcal{Y}_-$ via  the decomposition $E=V_+\oplus V_-$ and write the metric in terms of the bilinear form from \eqref{eq6} as
\begin{eqnarray}\label{eq65}
\mathcal{G}(\mathcal{A}_{\mathcal{X}}\mathcal{Y}, \mathcal{Z})&=&\mathcal{G}([\mathcal{X}, \mathcal{Y}]_C, \mathcal{Z})+\mathcal{G}(d<\mathcal{X}, \mathcal{Y}>, \mathcal{Z})-\mathcal{G}(\mathbb{D}^H_{\mathcal{X}}\mathcal{Y}, \mathcal{Z})\\
&=&<([\mathcal{X}, \mathcal{Y}]_C)_+, \mathcal{Z}_+>_++<(d<\mathcal{X}, \mathcal{Y}>)_+, \mathcal{Z}_+>_+-<(\mathbb{D}^H_{\mathcal{X}}\mathcal{Y})_+, \mathcal{Z}_+>_+\nonumber\\
&&-<([\mathcal{X}, \mathcal{Y}]_C)_-, \mathcal{Z}_->_--<(d<\mathcal{X}, \mathcal{Y}>)_-, \mathcal{Z}_->_-+<(\mathbb{D}^H_{\mathcal{X}}\mathcal{Y})_-, \mathcal{Z}_->_-.
\end{eqnarray}
By using torsion-free condition, we can write the Courant bracket $[\mathcal{X}, \mathcal{Y}]_C$ in terms of the generalized connection $\mathbb{D}^H$ and we have
\begin{eqnarray}\label{eq67}
\mathcal{G}(\mathcal{A}_{\mathcal{X}}\mathcal{Y}, \mathcal{Z})&=&<(\mathbb{D}^H_{\mathcal{X}}\mathcal{Y})_+, \mathcal{Z}_+>_+-<(\mathbb{D}^H_{\mathcal{Y}}\mathcal{X})_+, \mathcal{Z}_+>_+\nonumber\\
&&+\frac{1}{2}\bigg(<(\mathbb{D}^H_{\mathcal{Z}}\mathcal{X})_+, \mathcal{Y}_+>_+-<(\mathbb{D}^H_{\mathcal{Z}}\mathcal{Y})_+, \mathcal{X}_+>_+\bigg)\nonumber\\
&&+<(d<\mathcal{X}, \mathcal{Y}>)_+, \mathcal{Z}_+>_+-<(\mathbb{D}^H_{\mathcal{X}}\mathcal{Y})_+, \mathcal{Z}_+>_+\nonumber\\
&&-<(\mathbb{D}^H_{\mathcal{X}}\mathcal{Y})_-, \mathcal{Z}_->_-+<(\mathbb{D}^H_{\mathcal{Y}}\mathcal{X})_-, \mathcal{Z}_->_-\nonumber\\
&&-\frac{1}{2}\bigg(<(\mathbb{D}^H_{\mathcal{Z}}\mathcal{X})_-, \mathcal{Y}_->_--<(\mathbb{D}^H_{\mathcal{Z}}\mathcal{Y})_-, \mathcal{X}_->_-\bigg)\nonumber\\
&&-<(d<\mathcal{X}, \mathcal{Y}>)_-, \mathcal{Z}_->_-+<(\mathbb{D}^H_{\mathcal{X}}\mathcal{Y})_-, \mathcal{Z}_->_-\\
&=&-<(\mathbb{D}^H_{\mathcal{Y}}\mathcal{X})_+, \mathcal{Z}_+>_++<(d<\mathcal{X}, \mathcal{Y}>)_+, \mathcal{Z}_+>_+\nonumber\\
&&+\frac{1}{2}\bigg(<(\mathbb{D}^H_{\mathcal{Z}}\mathcal{X})_+, \mathcal{Y}_+>_+-<(\mathbb{D}^H_{\mathcal{Z}}\mathcal{Y})_+, \mathcal{X}_+>_+\bigg)\nonumber\\
&&+<(\mathbb{D}^H_{\mathcal{Y}}\mathcal{X})_-, \mathcal{Z}_->_--<(d<\mathcal{X}, \mathcal{Y}>)_-, \mathcal{Z}_->_-\nonumber\\
&&-\frac{1}{2}\bigg(<(\mathbb{D}^H_{\mathcal{Z}}\mathcal{X})_-, \mathcal{Y}_->_--<(\mathbb{D}^H_{\mathcal{Z}}\mathcal{Y})_-, \mathcal{X}_->_-\bigg) \nonumber \\ \label{eq68}
\end{eqnarray}
and similarly
\begin{eqnarray}\label{eq69}
\mathcal{G}(\mathcal{Y}, \mathcal{A}_{\mathcal{X}}\mathcal{Z})&=&-<\mathcal{Y}_+, (\mathbb{D}^H_{\mathcal{Z}}\mathcal{X})_+>_++<\mathcal{Y}_+, (d<\mathcal{X}, \mathcal{Z}>)_+>_+\nonumber\\
&&+\frac{1}{2}\bigg(<(\mathbb{D}^H_{\mathcal{Y}}\mathcal{X})_+, \mathcal{Z}_+>_+-<(\mathbb{D}^H_{\mathcal{Y}}\mathcal{Z})_+, \mathcal{X}_+>_+\bigg)\nonumber\\
&&+<\mathcal{Y}_-, (\mathbb{D}^H_{\mathcal{Z}}\mathcal{X})_->_--<\mathcal{Y}_-, (d<\mathcal{X}, \mathcal{Z}>)_->_-\nonumber\\
&&-\frac{1}{2}\bigg(<(\mathbb{D}^H_{\mathcal{Y}}\mathcal{X})_-, \mathcal{Z}_->_--<(\mathbb{D}^H_{\mathcal{Y}}\mathcal{Z})_-, \mathcal{X}_->_-\bigg).\nonumber\\
\end{eqnarray}
Then, the sum of \eqref{eq68} and \eqref{eq69} gives
\begin{eqnarray}\label{eq70}
\mathcal{G}(\mathcal{A}_{\mathcal{X}}\mathcal{Y}, \mathcal{Z})+\mathcal{G}(\mathcal{Y}, \mathcal{A}_{\mathcal{X}}\mathcal{Z})&=&-\frac{1}{2}\bigg((<\mathbb{D}^H_{\mathcal{Y}}\mathcal{X})_+, \mathcal{Z}_+>_++<(\mathbb{D}^H_{\mathcal{Z}}\mathcal{X})_+, \mathcal{Y}_+>_+\bigg)\nonumber\\
&&-\frac{1}{2}<(\mathbb{D}^H_{\mathcal{Z}}\mathcal{Y}+\mathbb{D}^H_{\mathcal{Y}}\mathcal{Z})_+, \mathcal{X}_+>_+\nonumber\\
&&+<(d<\mathcal{X}, \mathcal{Y}>)_+, \mathcal{Z}_+>_++<\mathcal{Y}_+, (d<\mathcal{X}, \mathcal{Z}>)_+>_+\nonumber\\
&&+\frac{1}{2}\bigg((<\mathbb{D}^H_{\mathcal{Y}}\mathcal{X})_-, \mathcal{Z}_->_-+<(\mathbb{D}^H_{\mathcal{Z}}\mathcal{X})_-, \mathcal{Y}_->_-\bigg)\nonumber\\
&&+\frac{1}{2}<(\mathbb{D}^H_{\mathcal{Z}}\mathcal{Y}+\mathbb{D}^H_{\mathcal{Y}}\mathcal{Z})_-, \mathcal{X}_->_-\nonumber\\
&&-<(d<\mathcal{X}, \mathcal{Y}>)_-, \mathcal{Z}_->_--<\mathcal{Y}_-, (d<\mathcal{X}, \mathcal{Z}>)_->_-
\end{eqnarray}
or we can write the last two terms as
\begin{equation}\label{eq71}
<d<\mathcal{X}, \mathcal{Y}>, \mathcal{Z}>=i_{\mathcal{Z}}d<\mathcal{X}, \mathcal{Y}>=\mathbb{D}^H_{\mathcal{Z}}<\mathcal{X}, \mathcal{Y}>=<\mathbb{D}^H_{\mathcal{Z}}\mathcal{X}, \mathcal{Y}>+<\mathcal{X}, \mathbb{D}^H_{\mathcal{Z}}\mathcal{Y}>
\end{equation}
\begin{equation}\label{eq72}
<\mathcal{Y}, d<\mathcal{X}, \mathcal{Z}>>=i_{\mathcal{Y}}d<\mathcal{X}, \mathcal{Z}>=\mathbb{D}^H_{\mathcal{Y}}<\mathcal{X}, \mathcal{Z}>=<\mathbb{D}^H_{\mathcal{Y}}\mathcal{X}, \mathcal{Z}>+<\mathcal{X}, \mathbb{D}^H_{\mathcal{Y}}\mathcal{Z}>.
\end{equation}
These identities can be obtained by using \eqref{bilinear1}, considering the action of generalized connection to a function $f$ in \eqref{eq57}, using the identity $\pi(\mathcal{X})f=i_{\mathcal{X}}df$ for a generalized vector $\mathcal{X}$ and considering the metric compatibility of $\mathbb{D}^H$. So, we have
\begin{eqnarray}\label{eq73}
\mathcal{G}(\mathcal{A}_{\mathcal{X}}\mathcal{Y}, \mathcal{Z})+\mathcal{G}(\mathcal{Y}, \mathcal{A}_{\mathcal{X}}\mathcal{Z})&=&-\frac{1}{2}\bigg(<(\mathbb{D}^H_{\mathcal{Y}}\mathcal{X})_+, \mathcal{Z}_+>_++<(\mathbb{D}^H_{\mathcal{Z}}\mathcal{X})_+, \mathcal{Y}_+>_+\bigg)\nonumber\\
&&-\frac{1}{2}\bigg(<(\mathbb{D}^H_{\mathcal{Z}}\mathcal{Y})_+, \mathcal{X}_+>_++<(\mathbb{D}^H_{\mathcal{Y}}\mathcal{Z})_+, \mathcal{X}_+>_+\bigg)\nonumber\\
&&+<(\mathbb{D}^H_{\mathcal{Z}}\mathcal{X})_+, \mathcal{Y}_+>_++<(\mathbb{D}^H_{\mathcal{Y}}\mathcal{X})_+, \mathcal{Z}_+>_+\nonumber\\
&&+<\mathcal{X}_+, (\mathbb{D}^H_{\mathcal{Z}}\mathcal{Y})_+>_++<\mathcal{X}_+, (\mathbb{D}^H_{\mathcal{Y}}\mathcal{Z})_+>_+\nonumber\\
&&+\frac{1}{2}\bigg(<(\mathbb{D}^H_{\mathcal{Y}}\mathcal{X})_-, \mathcal{Z}_->_-+<(\mathbb{D}^H_{\mathcal{Z}}\mathcal{X})_-, \mathcal{Y}_->_-\bigg)\nonumber\\
&&+\frac{1}{2}\bigg(<(\mathbb{D}^H_{\mathcal{Z}}\mathcal{Y})_-, \mathcal{X}_->_-+<(\mathbb{D}^H_{\mathcal{Y}}\mathcal{Z})_-, \mathcal{X}_->_-\bigg)\nonumber\\
&&-<(\mathbb{D}^H_{\mathcal{Z}}\mathcal{X})_-, \mathcal{Y}_->_--<(\mathbb{D}^H_{\mathcal{Y}}\mathcal{X})_-, \mathcal{Z}_->_-\nonumber\\
&&-<\mathcal{X}_-, (\mathbb{D}^H_{\mathcal{Z}}\mathcal{Y})_->_--<\mathcal{X}_-, (\mathbb{D}^H_{\mathcal{Y}}\mathcal{Z})_->_-\nonumber\\
&=&\frac{1}{2}\bigg(<(\mathbb{D}^H_{\mathcal{Y}}\mathcal{X})_+, \mathcal{Z}_+>_++<(\mathbb{D}^H_{\mathcal{Z}}\mathcal{X})_+, \mathcal{Y}_+>_+\bigg)\nonumber\\
&&+\frac{1}{2}\bigg(<(\mathbb{D}^H_{\mathcal{Z}}\mathcal{Y})_+, \mathcal{X}_+>_++<(\mathbb{D}^H_{\mathcal{Y}}\mathcal{Z})_+, \mathcal{X}_+>_+\bigg)\nonumber\\
&&-\frac{1}{2}\bigg(<(\mathbb{D}^H_{\mathcal{Y}}\mathcal{X})_-, \mathcal{Z}_->_-+<(\mathbb{D}^H_{\mathcal{Z}}\mathcal{X})_-, \mathcal{Y}_->_-\bigg)\nonumber\\
&&-\frac{1}{2}\bigg(<(\mathbb{D}^H_{\mathcal{Z}}\mathcal{Y})_-, \mathcal{X}_->_-+<(\mathbb{D}^H_{\mathcal{Y}}\mathcal{Z})_-, \mathcal{X}_->_-\bigg)\nonumber\\
&=&-(\mathbb{L}_{\mathcal{X}}\mathcal{G})(\mathcal{Y}, \mathcal{Z})
\end{eqnarray}
where we write the last equality from \eqref{eq60}. On the other hand, we can define the generalized exterior derivative
\begin{equation}\label{eq74}
d^{H}:=\mathcal{E}^A\wedge \mathbb{D}^H_{\mathcal{X}_A}
\end{equation}
and we can calculate
\begin{eqnarray}\label{eq75}
i_{\mathcal{X}}d^{H}\widetilde{\mathcal{Y}}&=&i_{\mathcal{X}}(\mathcal{E}^A\wedge \mathbb{D}^H_{\mathcal{X}_A}\widetilde{\mathcal{Y}})\nonumber\\
&=&i_{\mathcal{X}}\mathcal{E}^A\wedge \mathbb{D}^H_{\mathcal{X}_A}\widetilde{\mathcal{Y}}-\mathcal{E}^A\wedge i_{\mathcal{X}}\mathbb{D}^H_{\mathcal{X}_A}\widetilde{\mathcal{Y}}\nonumber\\
&=&\mathbb{D}^H_{\mathcal{X}}\widetilde{\mathcal{Y}}-\mathcal{E}^A\wedge\bigg(i_{\mathcal{X}}\mathbb{D}^H_{\mathcal{X}_A}\widetilde{\mathcal{Y}}+i_{\mathcal{X}_A}\mathbb{D}^H_{\mathcal{X}}\widetilde{\mathcal{Y}}\bigg)+\mathbb{D}^H_{\mathcal{X}}\widetilde{\mathcal{Y}}
\end{eqnarray}
where we have used the identities $(i_{\mathcal{X}}\mathcal{E}^A)\mathcal{X}_A=\mathcal{X}$ and $\mathcal{E}^A\wedge i_{\mathcal{X}_A}\mathbb{D}^H_{\mathcal{X}}\widetilde{\mathcal{Y}}=\mathbb{D}^H_{\mathcal{X}}\widetilde{\mathcal{Y}}$ since $\mathcal{E}^A\wedge i_{\mathcal{X}_A}\alpha=p\alpha$ for $\alpha$ a generalized $p$-form. So, we obtain
\begin{eqnarray}\label{eq76}
\mathbb{D}^H_{\mathcal{X}}\widetilde{\mathcal{Y}}&=&\frac{1}{2}i_{\mathcal{X}}d^{H}\widetilde{\mathcal{Y}}+\frac{1}{2}\mathcal{E}^A\wedge\bigg(i_{\mathcal{X}}\mathbb{D}^H_{\mathcal{X}_A}\widetilde{\mathcal{Y}}+i_{\mathcal{X}_A}\mathbb{D}^H_{\mathcal{X}}\widetilde{\mathcal{Y}}\bigg)\nonumber\\
&=&\frac{1}{2}i_{\mathcal{X}}d^{H}\widetilde{\mathcal{Y}}+\frac{1}{2}\mathcal{E}^A\wedge\bigg(\mathcal{G}(\mathbb{D}^H_{\mathcal{X}_A}\mathcal{Y}, \mathcal{X})+\mathcal{G}(\mathbb{D}^H_{\mathcal{X}}\mathcal{Y}, \mathcal{X}_A)\bigg)
\end{eqnarray}
and by using the previous calculations
\begin{equation}\label{eq77}
\mathbb{D}^H_{\mathcal{X}}\widetilde{\mathcal{Y}}=\frac{1}{2}i_{\mathcal{X}}d^{H}\widetilde{\mathcal{Y}}+\mathcal{E}^A\wedge\left(-(\mathbb{L}_{\mathcal{Y}}\mathcal{G})(\mathcal{X}, \mathcal{X}_A)-\frac{1}{2}\bigg(\mathcal{G}(\mathbb{D}^H_{\mathcal{X}_A}\mathcal{X}, \mathcal{Y})+\mathcal{G}(\mathbb{D}^H_{\mathcal{X}}\mathcal{X}_A, \mathcal{Y})\bigg)\right)
\end{equation}
where we have used the last equality in \eqref{eq73} which can be written as
\begin{equation}\label{eqn3}
-(\mathbb{L}_{\mathcal{Y}}\mathcal{G})(\mathcal{X},\mathcal{X}_A)=\frac{1}{2}\bigg(\mathcal{G}(\mathbb{D}^H_{\mathcal{X}}\mathcal{Y},\mathcal{X}_A)+\mathcal{G}(\mathbb{D}^H_{\mathcal{X}_A}\mathcal{Y},\mathcal{X})\bigg)+\frac{1}{2}\bigg(\mathcal{G}(\mathbb{D}^H_{\mathcal{X}_A}\mathcal{X},\mathcal{Y})+\mathcal{G}(\mathbb{D}^H_{\mathcal{X}}\mathcal{X}_A,\mathcal{Y})\bigg).
\end{equation}
If we choose $\mathcal{Y}$ as a generalized Killing vector $\mathcal{K}$ satisfying the generalized Killing equation
\[
\mathbb{L}_{\mathcal{K}}\mathcal{G}=0,
\]
we have
\begin{equation}\label{eq78}
\mathbb{D}^H_{\mathcal{X}}\widetilde{\mathcal{K}}=\frac{1}{2}i_{\mathcal{X}}d^{H}\widetilde{\mathcal{K}}-\frac{1}{2}\bigg(\mathcal{G}(\mathbb{D}^H_{\mathcal{X}_A}\mathcal{X}, \mathcal{K})+\mathcal{G}(\mathbb{D}^H_{\mathcal{X}}\mathcal{X}_A, \mathcal{K})\bigg)\mathcal{E}^A.
\end{equation}
Let us choose $\mathcal{X} = \mathcal{X}_B$ as a basis vector which is in the same direct sum component $V_{\pm}$ with $\mathcal{X}_A$, and consider the admissible metric with basis $\mathcal{X}_A=\{X_a+\epsilon e^a\}$ and $\mathcal{E}^A=\{e^a+\epsilon X_a\}$. Then, we have from \eqref{eq20}
\begin{equation}\label{eq79}
\mathbb{D}^H_{\mathcal{X}_A}\mathcal{X}_B+\mathbb{D}^H_{\mathcal{X}_B}\mathcal{X}_A=\nabla_{X_a}\left(\begin{array}{cc}X_b\\ \epsilon e_b\end{array}\right)+\nabla_{X_b}\left(\begin{array}{cc}X_a\\ \epsilon e_a\end{array}\right)+\epsilon \frac{1}{6}\left(\begin{array}{cc}\widetilde{i_{X_b}i_{X_a}H}\\ i_{\epsilon X_b}i_{X_a}H\end{array}\right)+\epsilon \frac{1}{6}\left(\begin{array}{cc}\widetilde{i_{X_a}i_{X_b}H}\\ i_{\epsilon X_a}i_{X_b}H\end{array}\right)
\end{equation}
The last two terms cancel because of anti-symmetry and if we choose normal coordinates, the first two terms also cancel and we have
\begin{equation}\label{eq80}
\mathbb{D}^H_{\mathcal{X}_A}\mathcal{X}_B+\mathbb{D}^H_{\mathcal{X}_B}\mathcal{X}_A=0
\end{equation}
and we find
\begin{equation}\label{eq81}
\mathbb{D}^H_{\mathcal{X}_A}\widetilde{\mathcal{K}}=\frac{1}{2}i_{\mathcal{X}_A}d^{H}\widetilde{\mathcal{K}}
\end{equation}

with respect to an admissible basis $\mathcal{X}_A = \{X_a + \epsilon e^a\}$ which is in the same direct sum component $V_{\pm}$ with $\mathcal{K}$. So, \eqref{eq81} corresponds to the generalized Killing equation in terms of the generalized connection. Since $\mathcal{X}_A$ and $\mathcal{K}$ are in the same direct sum component $V_{\pm}$, the cross terms involving opposite direct sum components do not appear. Importantly, the equivalence of the left-hand side with the term involving the unique generalized exterior derivative $d^H$ ensures that the equation is well-defined, independent of the ambiguity in the specific choice of the generalized connection. As the anti-symmetric generalization of generalized Killing equation, we can write the generalized KY equation for generalized $p$-forms $\alpha$ which only includes the same direct sum components $V_{\pm}$ with $\mathcal{X}_A$ as

\begin{equation}\label{eq82}
\mathbb{D}^H_{\mathcal{X}_A}\mathcal{\alpha}=\frac{1}{p+1}i_{\mathcal{X}_A}d^{H}\mathcal{\alpha}
\end{equation}
where $d^H=\mathcal{E}^A\wedge\mathbb{D}^H_{\mathcal{X}_A}=d^G+\frac{\epsilon}{4}H\underset{1}{\wedge}$ where $d^G$ is defined as $d^G=\mathcal{E}^A\wedge\nabla_{\pi({\mathcal{X}_A})}$. Here $H\underset{1}{\wedge}=i_{X^a}H\wedge i_{X_a}$. By using \eqref{eq34}, \eqref{eq82} is written explicitly as
\begin{equation}\label{eq84}
\mathbb{D}^H_{\mathcal{X}_A}\mathcal{\alpha}=\nabla_{X_a}\mathcal{\alpha}-\epsilon\frac{1}{12}[i_{X_a}H,\mathcal{\alpha}]_{Cl}=\frac{1}{p+1}i_{X_a}\left(d^G\mathcal{\alpha}+\epsilon\frac{1}{4}H\underset{1}{\wedge}\mathcal{\alpha}\right) 
\end{equation}
in terms of flux terms which resembles the KY equation modified by 3-form flux $H$ corresponding to torsion. The restriction $X_A\in V_{\pm}$ is imposed to keep the condition compatible with the splitting
defined by the generalised metric and to avoid mixing between $V_{+}$ and $V_{-}$ components in
the defining equation. Importantly, this restriction should not be interpreted as implying
that a metric-compatible torsion-free generalised connection is unique.
Rather, the usefulness of \eqref{eq82} is that it admits an equivalent \emph{intrinsic} formulation
in terms of the induced torsionful metric connection and the $H$-twisted differential, making the
dependence on an arbitrary choice of generalised Levi--Civita connection drop out at the level of
the final generalised Killing--Yano condition; explicitly, \eqref{eq82} is equivalent to \eqref{eq84}.

\section{Relation Between Generalized KY Forms and Generalized Geometry Killing Spinors}

We will investigate the relation between generalized geometry Killing spinors satisfying \eqref{eq38} and generalized KY forms satisfying \eqref{eq82}. As noted previously, this construction utilizes the $SO(n,n)$ spinor representation. While distinct from the standard supersymmetry parameters in supergravity, the bilinears constructed from these generalized spinors offer a formal extension of the Killing-Yano structure to the generalized geometry framework. If we have two generalized geometry Killing spinors $\phi$ and $\psi$ satisfying \eqref{eq38}, we can construct generalized $p$-form bilinears $(\psi\overline{\phi})_p + (\phi\overline{\psi})_p$ where a $p$-form bilinear is written in terms of the generalized form basis as
\[
(\psi\overline{\phi})_p=(\phi, \mathcal{E}_{A_p...A_2A_1}.\psi)\mathcal{E}^{A_1A_2...A_p}.
\]
We can calculate the generalized derivative of $p$-form bilinears as
\begin{eqnarray}\label{eq85}
\mathbb{D}^H_{\mathcal{X}_A}\left[(\psi\overline{\phi})_p+(\phi\overline{\psi})_p\right]&=&((\mathbb{D}^H_{\mathcal{X}_A}\psi)\overline{\phi})_p+(\psi\overline{(\mathbb{D}^H_{\mathcal{X}_A}\phi)})_p+((\mathbb{D}^H_{\mathcal{X}_A}\phi)\overline{\psi})_p+(\phi\overline{(\mathbb{D}^H_{\mathcal{X}_A}\psi)})_p\nonumber\\
&=&(\lambda\mathcal{E}_A.\psi\overline{\phi})_p+(\psi\overline{(\lambda \mathcal{E}_A.\phi)})_p+(\lambda\mathcal{E}_A.\phi\overline{\psi})_p+(\phi\overline{(\lambda \mathcal{E}_A.\psi)})_p
\end{eqnarray}
from the generalized geometry Killing spinor equation. Here, we use the property $\overline{\mathbb{D}_{\mathcal{X}_A}\psi}=\mathbb{D}_{\mathcal{X}_A}\overline{\psi}$. By using the definition $\overline{\psi}\phi=(\psi,\phi)$, we can write for a generalized geometry spinor $\kappa$
\begin{equation}\label{eq86}
\psi(\overline{\lambda\mathcal{E}_A.\phi})\kappa=(\lambda\mathcal{E}_A.\phi, \kappa)\psi=\lambda^j(\phi, \mathcal{E}_A^{\mathcal{J}}.\kappa)\psi=\lambda^j(\psi\overline{\phi})\mathcal{E}_A^{\mathcal{J}}.\kappa
\end{equation}
and since $j$ is identity and $\mathcal{J}=\xi\eta$ for generalized geometry spinors, we have
\begin{eqnarray}\label{eq87}
\psi(\overline{\lambda\mathcal{E}_A.\phi})&=&\lambda^j(\psi\overline{\phi}).\mathcal{E}_A^{\mathcal{J}}=-\lambda(\psi\overline{\phi}).\mathcal{E}_A\\
\phi(\overline{\lambda\mathcal{E}_A.\psi})&=&\lambda^j(\phi\overline{\psi}).\mathcal{E}_A^{\mathcal{J}}=-\lambda(\phi\overline{\psi}).\mathcal{E}_A.\label{eq88}
\end{eqnarray}
Then, we obtain
\begin{equation}\label{eq89}
\mathbb{D}^H_{\mathcal{X}_A}\left[(\psi\overline{\phi})_p+(\phi\overline{\psi})_p\right]=(\lambda\mathcal{E}_A.\psi\overline{\phi})_p-(\lambda(\psi\overline{\phi}).\mathcal{E}_A)_p+(\lambda\mathcal{E}_A.\phi\overline{\psi})_p-(\lambda(\phi\overline{\psi}).\mathcal{E}_A)_p.
\end{equation}
For a generalized $p$-form $\mathcal{\alpha}$, we have
\begin{eqnarray}\label{eq90}
\mathcal{E}_A.\mathcal{\alpha}&=&i_{\mathcal{X}_A}\mathcal{\alpha}+\mathcal{\widetilde{E}}_A\wedge\mathcal{\alpha}\\
\mathcal{\alpha}.\mathcal{E}_A&=&-i_{\mathcal{X}_A}\eta\mathcal{\alpha}+\mathcal{\widetilde{E}}_A\wedge\eta\mathcal{\alpha}\label{eq91}
\end{eqnarray}
which are generalizations of the Clifford product equalities for ordinary differential forms. So, we can write \eqref{eq89} as
\begin{eqnarray}\label{eq92}
\mathbb{D}^H_{\mathcal{X}_A}\left[(\psi\overline{\phi})_p+(\phi\overline{\psi})_p\right]&=&\lambda \mathcal{E}_A\wedge(\psi\overline{\phi})_{p-1}+\lambda i_{\mathcal{X}_A}(\psi\overline{\phi})_{p+1}-\lambda \mathcal{E}_A\wedge(\psi\overline{\phi})^{\eta}_{p-1}+\lambda i_{\mathcal{X}_A}(\psi\overline{\phi})^{\eta}_{p+1}\nonumber\\
\ & +&\lambda \mathcal{E}_A\wedge(\phi\overline{\psi})_{p-1}+\lambda i_{\mathcal{X}_A}(\phi\overline{\psi})_{p+1}-\lambda \mathcal{E}_A\wedge(\phi\overline{\psi})^{\eta}_{p-1}+\lambda i_{\mathcal{X}_A}(\phi\overline{\psi})^{\eta}_{p+1}
\end{eqnarray}
where we extend the spinor bilinears as $\psi\overline{\phi}=(\psi\overline{\phi})_0+...+(\psi\overline{\phi})_p+(\psi\overline{\phi})_{p+1}+...$ in terms of component forms. From the definitions of generalized exterior and co-derivatives
\begin{eqnarray}\label{eq93}
d^H&=&\mathcal{E}^A\wedge\mathbb{D}^H_{\mathcal{X}_A}\\
\delta^H&=&-i_{\mathcal{X}^A}\mathbb{D}^H_{\mathcal{X}_A} \label{eq94}
\end{eqnarray}
we obtain
\begin{eqnarray}\label{eq95}
d^H\left[(\psi\overline{\phi})_p+(\phi\overline{\psi})_p\right]&=&\mathcal{E}^A\wedge\mathbb{D}_{\mathcal{X}_A}\left[(\psi\overline{\phi})_p+(\phi\overline{\psi})_p\right]\nonumber\\
&=&\lambda (p+1)(\psi\overline{\phi})_{p+1}+\lambda (p+1)(\psi\overline{\phi})_{p+1}^{\eta}\nonumber\\
&&+\lambda (p+1)(\phi\overline{\psi})_{p+1}+\lambda (p+1)(\phi\overline{\psi})_{p+1}^{\eta}
\end{eqnarray}
where we have used the identity $\mathcal{E}^A\wedge i_{\mathcal{X}_A}\alpha=p\alpha$ for a generalized $p$-form $\alpha$ and similarly
\begin{eqnarray}\label{eq96}
\delta^H\left[(\psi\overline{\phi})_p+(\phi\overline{\psi})_p\right]&=&-i_{\mathcal{X}^A}\mathbb{D}_{\mathcal{X}_A}\left[(\psi\overline{\phi})_p+(\phi\overline{\psi})_p\right]\nonumber\\
&=&-\lambda (2n-p+1)(\psi\overline{\phi})_{p-1}^{\eta}+\lambda (2n-p+1)(\psi\overline{\phi})_{p-1}^{\eta}\nonumber\\
&&-\lambda (2n-p+1)(\phi\overline{\psi})_{p-1}^{\eta}+\lambda (2n-p+1)(\phi\overline{\psi})_{p-1}^{\eta}.
\end{eqnarray}
Since $\lambda$ is real, we can consider the cases of $p$ is even or odd separately. For $p$ is odd, we have
\begin{eqnarray}\label{eq97}
\mathbb{D}^H_{\mathcal{X}_A}\left[(\psi\overline{\phi})_p+(\phi\overline{\psi})_p\right]&=&2\lambda i_{\mathcal{X}_A}\left[(\psi\overline{\phi})_{p+1}+(\phi\overline{\psi})_{p+1}\right]\\
d^H\left[(\psi\overline{\phi})_p+(\phi\overline{\psi})_p\right]&=&2\lambda (p+1)\left[(\psi\overline{\phi})_{p+1}+(\phi\overline{\psi})_{p+1}\right] \label{eq98}\\
\delta^H\left[(\psi\overline{\phi})_p+(\phi\overline{\psi})_p\right]&=&0 \label{eq99}
\end{eqnarray}
and by comparing them, we find
\begin{equation}\label{eq100}
\mathbb{D}^H_{\mathcal{X}_A}\left[(\psi\overline{\phi})_p+(\phi\overline{\psi})_p\right]=\frac{1}{p+1}i_{\mathcal{X}_A}d^G\left[(\psi\overline{\phi})_p+(\phi\overline{\psi})_p\right]
\end{equation}
which corresponds to the generalized KY equation. For $p$ is even, we have
\begin{eqnarray}\label{eq101}
\mathbb{D}^H_{\mathcal{X}_A}\left[(\psi\overline{\phi})_p+(\phi\overline{\psi})_p\right]&=&2\lambda\mathcal{E}_A\wedge\left[(\psi\overline{\phi})_{p-1}+(\phi\overline{\psi})_{p-1}\right]\\
d^H\left[(\psi\overline{\phi})_p+(\phi\overline{\psi})_p\right]&=&0\label{eq102}\\
\delta^H\left[(\psi\overline{\phi})_p+(\phi\overline{\psi})_p\right]&=&-2\lambda (2n-p+1)\left[(\psi\overline{\phi})_{p-1}+(\phi\overline{\psi})_{p-1}\right] \label{eq103}
\end{eqnarray}
and by comparing them, we find
\begin{equation}\label{eq104}
\mathbb{D}^H_{\mathcal{X}_A}\left[(\psi\overline{\phi})_p+(\phi\overline{\psi})_p\right]=-\frac{1}{2n-p+1}\mathcal{E}_A\wedge\delta^H\left[(\psi\overline{\phi})_p+(\phi\overline{\psi})_p\right].
\end{equation}
This equation corresponds to the generalized closed conformal KY equation written for a $p$-form $\omega$ as
\begin{equation}\label{eqn4}
\nabla_{X_a}\omega=-\frac{1}{n-p+1}e_a\wedge\delta\omega.
\end{equation}
Then, we show that generalized $p$-form bilinears constructed out of generalized geometry Killing spinors correspond to generalized KY forms when $p$ is odd and to generalized closed conformal KY forms when $p$ is even. In fact, generalized KY equation and generalized closed conformal KY equation are special cases of the generalized conformal KY equation which is written as
\begin{equation}\label{eq105}
\mathbb{D}^H_{\mathcal{X}_A}\mathcal{\alpha}=\frac{1}{p+1}i_{\mathcal{X}_A}d^H\mathcal{\alpha}-\frac{1}{2n-p+1}\mathcal{E}_A\wedge\delta^H\mathcal{\alpha}.
\end{equation}
Hence, if we have a sum of even and odd generalized $p$-form bilinears constructed out of generalized geometry Killing spinors, then they correspond to generalized conformal KY forms.

\section{Conclusion}

From the definition of spinor inner product for generalized geometry spinors, we find the properties of generalized spinor bilinears for different cases. By extending the concepts of geometric Killing spinors and twistor spinors to the framework of generalized geometry, we construct the equations corresponding to generalized geometry Killing and twistor spinors in terms of differential forms and especially we show that generalized Killing spinor equation corresponds to conformal KY equation with torsion. We extend the concept of KY forms to generalized geometry by proving that the generalized Killing equation written in terms of the generalized Lie derivative can also be described in terms of the generalized connection. Moreover, we show that bilinear forms of generalized Killing spinors with different parity correspond to generalized KY forms for the odd case and closed conformal KY forms for the even case and in general they correspond to conformal KY forms for mixed parity.

As in the case of KY forms and geometric Killing spinors which can be used in the construction of extended Killing superalgebra structures, the construction of generalized KY forms and their relation with generalized Killing spinors open the way to investigate the construction of superalgebra structures generated by generalized hidden symmetries. While the physical interpretation of these $SO(n,n)$ spinor bilinears differs from the standard bilinears of supergravity Killing spinors (which reside in the split bundle $V_{\pm}$), the mathematical parallels established here suggest that these generalized structures may provide a useful formal framework for analyzing hidden symmetries in generalized geometry.

\begin{acknowledgments}

This study was supported by Scientific and Technological Research Council of T\"urkiye (T\"UB\.ITAK) under the Grant Number 123F261. The authors thank T\"UB\.ITAK for their supports.

\end{acknowledgments}

%\references%

%\begin{references}

%\end{references}

\end{document}